\documentclass[12pt,preprint]{aastex}
\usepackage[]{natbib}
\usepackage{graphicx, amsmath}
\begin{document}
\title{Discovery of a Radio-Selected z $\sim$ 6 Quasar}
\author{Gregory R. Zeimann \altaffilmark{1}, Richard L. White \altaffilmark{2}, Robert H. Becker \altaffilmark{1,3}, Jacqueline A. Hodge \altaffilmark{1,4}, Spencer A. Stanford \altaffilmark{1,3}, Gordon T. Richards \altaffilmark{5}}
\altaffiltext{1}{Department of Physics, University of California, Davis, CA 95616}
\altaffiltext{2}{Space Telescope Science Institute, Baltimore, MD 21218}
\altaffiltext{3}{IGPP/Lawrence Livermore National Laboratory}
\altaffiltext{4}{Max-Planck-Institut f\"{u}r Astronomie, K\"{o}nigstuhl 17, D-69117 Heidelberg, Germany}
\altaffiltext{5}{Department of Physics, Drexel University, 3141 Chestnut Street, Philadelphia, PA 19104}

\section{Abstract}

We present the discovery of only the second radio-selected, z $\sim$ 6 quasar.  We identified SDSS J222843.54+011032.2 (z=5.95) by matching the optical detections of the deep Sloan Digital Sky Survey (SDSS) Stripe 82 with their radio counterparts in the Stripe82 VLA Survey.  We also matched the Canadian-France-Hawaiian Telescope Legacy Survey Wide (CFHTLS Wide) with the Faint Images of the Radio Sky at Twenty-cm (FIRST) survey but have yet to find any z $\sim$ 6 quasars in this survey area.  The discovered quasar is optically-faint, $z = 22.3$ and M$_{1450}$ $\sim$ -24.5, but radio-bright, with a flux density of f$_{1.4GHz, peak}$ = 0.31mJy and a radio-loudness of R $\sim$ 1100 (where R $\equiv$ $f_{5GHz}/f_{2500}$).  The $i-z$ color of the discovered quasar places it outside the color selection criteria for existing optical surveys.  We conclude by discussing the need for deeper wide-area radio surveys in the context of high-redshift quasars.    
 
\section{Introduction}
\label{sec:intro}

High-redshift quasars (z $\sim$ 6) which are powered by supermassive black holes (SMBH's) provide a window into the early universe, and, unsurprisingly, the study of these rare objects has grown in the decade since their discovery (\citealp{2000AJ....120.1167F}; \citealp{2001AJ....122.2833F}).  These distant SMBH's are important in the study of galaxy evolution and the epoch of re-ionization.   The strong correlation observed between the velocity dispersion of the stars in a galaxy and the mass of its central SMBH indicates a linkage between the mass of the SMBH and the evolution of the galaxy \citep{2000ApJ...539L...9F}.  This relation may be explained by a ``quasar phase" of the SMBH, where significant accretion is coupled with feedback into the stellar formation processes of the host galaxy \citep{ 2005ApJ...625L..71H}.  Since nearly every galaxy is thought to have a central SMBH, active or not, high-redshift quasars are an important piece of the puzzle in understanding early galaxy evolution.  

The spectra of distant quasars probe the intergalactic medium (IGM) and can be used to study the re-ionization epoch along the line of sight (\citealp{2001AJ....122.2850B}; \citealp{2003AJ....126....1W}; \citealp{2006AJ....132..117F}).  The presence of the Gunn-Peterson trough \citep{1965ApJ...142.1633G} can indicate the neutral hydrogen fraction and act as a flag marking the re-ionization epoch.  There is evidence that the fraction of neutral hydrogen by volume increases an order of magnitude from z $=$ 5.7 to z $=$ 6.4 (\citealp{2006AJ....132..117F}; \citealp{2010ApJ...714..834C}).  This increase may mark the end of the re-ionization epoch, placing a high importance on any new discoveries of z $>$ 5.7 quasars, especially those which are radio-loud.  Radio-loud quasars are valuable as probes of the early IGM because the coming generation of radio telescopes (EVLA, ALMA, SKA, LOFAR, etc.) will use them to measure the properties of neutral hydrogen through 21 cm absorption studies (\citealp{2004NewAR..48.1029C}; \citealp{2010HiA....15..312K}).

Despite large area surveys designed to find high-redshift quasars, only about 60 ($z > 5.7$) quasars have been found so far (\citealp{2000AJ....120.1167F}; \citealp{2007AJ....134.2435W}; \citealp{2008AJ....135.1057J}).  These surveys all find quasars using similar methods that rely on red optical colors ($(i - z)_{AB}$ $>$ 2) and blue near-IR colors ($(z - J)_{AB}$ $<$ 1).  The red optical color of high-z quasars is a consequence of the Ly$\alpha$ break moving through and out of the blueward optical bands.  The blue near-IR color cut is used to distinguish high-redshift quasars from the largest contaminant in these surveys: cool dwarf stars.  An alternative method to separate high-redshift quasars from cool dwarf stars is to require a radio detection, as very few stars are radio-bright (\citealp{kimball}).  This method only finds $\sim$5$\%$ of quasars (3 of $\sim$60 z $\sim$ 6 quasars have been detected at $>$1mJy in the radio), but it can be used to select quasars with colors that fail the usual optical/near-IR selection.  Previously, one radio-selected z $\sim$ 6 quasar was found by \citet{2006ApJ...652..157M} in a mere 4 ${deg}^2$ search area in the NOAO Deep Wide-Field Survey (NDWFS).  This quasar was bright enough to be discovered by other surveys; however, its red near-IR colors placed it in the color space occupied by cool dwarf stars, and hence it was missed by typical color-color selection methods.  

This work presents the discovery of the second radio-selected, z $\sim$ 6 quasar.  We used two different survey combinations to search for radio-selected z $>$ 5.7 quasars: the Canadian-France-Hawaiian Telescope Legacy Survey Wide (CFHTLS Wide \footnote[6]{http://www.cfht.hawaii.edu/Science/CFHTLS/}) matched to the Faint Images of the Radio Sky at Twenty-cm (FIRST - \citealp{1995ApJ...450..559B}) survey, and the deep Sloan Digital Sky Survey (SDSS Stripe 82 - \citealp{2009ApJS..182..543A}) matched with the Stripe82 VLA Survey (Hodge et al. 2011).  We discovered the quasar in Stripe 82 with the deeper radio data of the Stripe82 VLA Survey.  

 In \S 3 we describe our candidate selection methods and expected number of discoveries.  In \S 4, we discuss the observations of SDSS J222843.53+011032.0 (SDSS J2228+0110; z $=$ 5.95). In \S 5 we discuss the implications of further studies of high redshift radio-selected quasars.  All magnitudes are AB unless stated otherwise. This paper assumes a flat cosmological model with $\Omega_{m} = 0.28$, $\Omega_{\Lambda} = 0.72$, and $H_{0} = 70$ km s$^{-1}$  Mpc$^{-1}$. 

\section{Candidate Selection}

\subsection{CFHTLS Wide}

CFHTLS Wide is an intermediate depth and area survey covering 171 deg$^2$ with $\sim$130 deg$^2$ publicly available at the time of this work (release T0005).  The typical integration times are 4300s in $i$ and 3600s in $z$, reaching an average depth of 24.5 in i and 23.8 in z at the 80$\%$ completeness limit determined through simulation\footnote[7]{http://terapix.iap.fr/cplt/oldSite/Descart/CFHTLS-T0005-Release.pdf}. 

\subsection{FIRST}

FIRST is a 20 cm survey over $\sim$10,000 deg$^2$.  With a sensitivity threshold of 1 mJy, it achieves a source density of $\sim$90 deg$^{-2}$.  The FIRST survey covers the entirety of the CFHTLS Wide survey, and the typical astrometric accuracy between the two surveys is $<$ 0.5''.       

\subsection{Stripe 82}

During the months when the primary SDSS area was not observable, SDSS repeatedly observed a strip of sky along the Galactic Equator known as Stripe 82.  This patch of sky is 300 deg$^2$ and spans roughly 20$^{h}$ $<$ RA $<$ 4$^{h}$ and -1.5$^{\circ}$ $<$ Dec $<$ 1.5$^{\circ}$.  The resulting co-additions (\citealp{2009ApJS..182..543A}) go two magnitudes deeper than the typical SDSS images and reach 23.3 in $i$ and 22.5 in $z$ at the 95$\%$ repeatable detection limit.  

\subsection{Stripe 82 VLA Survey}

This 1.4 GHz survey was conducted with the Very Large Array (VLA) in A-configuration and has an angular resolution of 1.8'' (Program ID AR646 and AR685).  It achieves a median rms noise of 52 $\mu$Jy beam$^{-1}$ over 92 deg$^{2}$ (\citealp{hodge2011}), making it the deepest 1.4 GHz survey to cover that much sky.  A catalog of 17,969 isolated radio components, for an overall source density of $\sim$195 sources deg$^{-2}$, is publicly available.  The astrometric accuracy of the data is excellent, with an rms scatter of 0.25'' in both right ascension and declination when matched to SDSS's Stripe 82.

\subsection{Selection Method}

High-redshift quasars are typically selected based on their very red optical colors: $(i - z)_{AB} > 1.5$.  Cool dwarf stars also have red optical colors, so in an effort to reduce contamination, surveys such as SDSS require a blue near-IR color (see Figure 1).  The optical color probes the Ly$\alpha$ break while the near-IR color probes the quasar continuum as well as strong emission features.  Dust reddening can affect the color of the quasar continuum and displace a fraction of the high-z quasars outside of the typical selection criterion.  An alternative to a blue near-IR color cut is radio-selection, which reduces the contamination from cool dwarf stars to nearly zero but will only be able to recover $\sim$5\% of high-z quasars.   

We combined optical and radio data through catalog matching.  Counterparts between catalogs were defined using a matching radius of 0.6'' for Stripe 82 and 1'' for CFHTLS Wide, which was a  compromise between completeness and reliability.  We used a method similar to that of \citet{2002ApJS..143....1M} to estimate completeness and reliability of the matches as a function of radius.  A matching radius of 1'' for FIRST-CFHTLS wide results in a completeness of 83\% and a reliability of 94\%.  A matching radius of 0.6'' for Stripe 82-Stripe 82 VLA results in a completeness of 91\% and a reliability of 99\%.   If there were multiple optical sources for a single radio source, then only the closest match was used.   After matching the two catalogs, two photometric cuts were applied to select the initial $z > 6$ quasar candidates:  $(i -z)_{AB} > 1.5$ and $z_{AB} <  23.8$ for CFHTLS Wide, and $(i -z)_{AB} > 1.7$ \footnote[8]{The color conversion between SDSS and CFHT for typical z $\sim$ 6 quasars is $(i-z)_{SDSS} > (i-z)_{CFHT} + 0.20$;  http://www.cadc.hia.nrc.gc.ca/megapipe/docs/filters.html} and $z_{AB} <  22.5$ for Stripe 82.  The candidates went through another series of cuts which required that the \textit{u}, \textit{g}, and \textit{r} band fluxes be below the 3$\sigma$ detection limit.  Also, a visual inspection of the \textit{i} and \textit{z} band images was conducted to ensure there were no cosmic rays or bad pixels contaminating the photometry.  The remaining candidates were then checked against known sources as to not repeat observations.  One of the candidates was a previously found z$=$6.21 quasar by \citet{2010AJ....139..906W}, CFHQS J1429+5447, and is the third radio-loud, z $\sim$ 6 quasar to be identified.  CFHQS J1429+5447 is the radio-brightest z $\sim$ 6 quasar yet to be found with f$_{1.4GHz, peak}$ = 2.93 mJy and has the highest radio-loudness value of R $\sim$ 3200, where R $\equiv$ $f_{5 GHz} / f_{2500\AA}$.  After all of the cuts, there were 29 remaining candidates in CFHTLS Wide, i.e. $\sim$ 0.2 per deg$^2$, and 27 in Stripe 82, giving $\sim$ 0.3 per deg$^2$.  

Our search area was previously mined by other high-z quasar searches and thus our candidate list has some overlap with those selection methods.  In the case of CFHTLS Wide, 3 of our 29 candidates satisfy the criteria set forth by \citet{2009AJ....137.3541W}, $(i -z)_{AB} > 2.0$ and $z_{AB} < 23.0$ or 10$\sigma$ $z_{AB}$ limit for the field (private communication).  Only 3 of the 27 candidates in Stripe 82 would have been selected by \citet{2009AJ....138..305J} while the other 24 candidates were either too blue, $(i -z)_{AB} < 2.2$, or too faint, $z_{AB} > 21.8$ for their completeness-limited selection.      

To properly estimate how many quasars our radio-selection should find, we had to first estimate the completeness of our method as a function of redshift and rest-frame absolute magnitude, $M_{1450}$\footnote[9]{$M_{1450}$ was calculated using the measured z-band magnitudes and assuming the \citet{2001AJ....122..549V} spectrum corrected for the effective Gunn-Peterson optical depth due to Ly$\alpha$ and Ly$\beta$ absorption (\citealp{2006AJ....132..117F})}.  The completeness of our optical selection was estimated through simulation. A relation between redshift and $i-z$ color was calculated using the median color track of z $\sim$ 3 quasars from SDSS redshifted from z$=$5.5 to z$=$6.7 and corrected for the effective Gunn-Peterson optical depth due to Ly$\alpha$ and Ly$\beta$ absorption (\citealp{2006AJ....132..117F}).  This track is very similar to that calculated by \citet{2009AJ....137.3541W} as shown in Figure 1.  A ``measured'' $i-z$ color was drawn randomly from a gaussian distribution with a mean $i-z$ color taken from the median quasar color track and a sigma calculated using both the standard deviation of the median quasar color track and the median error as a function of the $i$ and $z$ magnitudes.  This was done hundreds of times for a grid of (z, $M_{1450}$), and the completeness was estimated by the fraction recovered by our selection method.  An example of our completeness as a function of redshift and absolute magnitude is shown in Figure 2.  The simulation of completeness only took into account the optical selection method and not the radio.  The radio-loud ($>$1mJy) fraction of quasars at 0 $<$ z $<$ 5 in SDSS is $\sim$10\%, but is strongly dependent on redshift and optical luminosity (\citealp{jiang2007}).  We adopt a radio-loud fraction of 5\% for z $\sim$ 6 and apply this fraction to estimate the number of radio-selected quasars expected.                

Using the luminosity function for z $=$ 6 quasars calculated by \citet{2010AJ....139..906W}, we are able to estimate the expected number of quasars in our search, N, \begin{equation}
N = A * \iint \ \Phi(z,M_{1450}) \, V_{c}(z)  \,\, p(z,M_{1450})   \, \mathrm{d}z \, \mathrm{d}M_{1450}.  
\end{equation}
The comoving volume, $V_{c}(z)$,  is corrected for the completeness, $p(z,M_{1450})$, to form an effective volume.  The luminosity function, $\Phi(z,M_{1450})$, is in the form of a double power law, and we used the best fit parameters from \citet{2010AJ....139..906W}.  Our search area is represented by A, in units of steradians.  From the calculation,  we expect 0.7 radio-selected quasars in Stripe 82 and 3.2 in CFHTLS Wide.  We find more candidates than the expected number of quasars for several reasons.  Firstly, some of our candidates are expected to be lower redshift, radio-loud AGN that share a red $i-z$ color with high-z quasars due to a strong Balmer break or dust extincted continuum.  Secondly, a few of our candidates may be spurious matches where the optical source is not really associated with the radio emission.  Thirdly, the use of a constant radio-loud fraction may not be appropriate as it has been observed to be a function of optical luminosity which could change our number estimates as much as a factor of 2 (\citealp{jiang2007}).    Lastly, the best fit parameters of the luminosity function from \citet{2010AJ....139..906W} have large errors and can change the expected number of quasars by a factor of 3 or 4.           

\section{Observations}

\subsection{Optical and Radio}

The quasar, SDSS J2228+0110, was selected for followup using the point spread function (PSF) magnitudes from the  co-added imaging catalog of Stripe 82 (\citealp{2009ApJS..182..543A}).  It was found near our $z$-band magnitude limit at $z$ $=$ 22.28 and near our $i-z$ color limit at $i-z$ $=$ 1.81 which would have been too faint and too blue to be selected by \citet{2009AJ....138..305J}.  It is one of the faintest quasars found to date, with M$_{1450}$ $=$ -24.53.  The quasar overlaps the UKIRT Infrared Deep Sky Survey (UKIDSS) sky coverage, but is undetected in the $J$-band placing an upper limit at $z-J \le 1.4$\footnote[10]{UKIDSS Large Area Survey detection limit is $J_{AB} = 20.9$, http://www.ukidss.org/surveys/surveys.html}.

The quasar was detected in the radio in the Stripe82 VLA Survey, which has a detection limit of 0.30 mJy.  SDSS J2228+0110 has a measured peak flux density of 0.31 mJy, just above the detection limit, and it is only the fourth z $\sim$ 6 quasar discovered with a flux density f$_{1.4GHz} >$  0.3 mJy.  The faint optical luminosity and relatively high radio luminosity, $L_{5 GHz} = $ 2.55 $\times$ 10$^{32}$, makes this one of the most radio-loud z $\sim$ 6 quasars ever found, with a radio-loudness of R $\sim$ 1100.  There is no discernible morphology from the optical or radio images given that the discovered quasar was at the limits of detection in both wavelengths (see Figure 3).

\subsection{Quasar Spectrum}

From thirty-five candidates observed in June and December 2010 (sixteen from CFHTLS Wide and nineteen from Stripe 82 of which most remain unidentified with featureless continua), we discovered one quasar at z=5.95, SDSS J2228+0110 (see Table 1).  The discovery spectrum of SDSS J2228+0110 was taken using the Keck I telescope with the Low Resolution Imaging Spectrometer (LRIS; \citealp{1995PASP..107..375O}).  The spectrum included four exposures of 900 seconds each and a 1" long slit.  It was taken with a 600/10000 grating on the red camera, resulting in a dispersion of $\sim$0.8$\AA$ per pixel.  The conditions were fair when the spectrum was taken, with $\sim$1'' seeing at a high air mass of 1.5.  

The reduction of the spectrum was done in a standard way.  Bias frames were obtained and combined with the overscan bias, then subtracted off the science image.  A flat field correction was applied using flat field frames taken during the observing run.  A specialized routine was used for cosmic ray rejection which recognized and rejected cosmic rays based on their shape.  The spectrum of our quasar was extracted using optimal variance weighting through the IRAF task apall.  The wavelength was calibrated using arc line lamps.  The standard star used for flux calibration was Feige 34\footnote[11]{The calibration flux was obtained from the Space Telescope standard star flux catalog} in combination with a custom Mauna Kea extinction curve as a function of wavelength.  A lower resolution spectrum, $\sim$8$\AA$ per pixel, was created using inverse sky-variance weighting.  The spectrum clearly identifies the source as a z=5.95 quasar with a large continuum break blueward of a strong emission line marked as Ly$\alpha$ (see Figure 4).

The only strong emission feature in the discovery spectrum is Ly$\alpha$.  The noise from sky emission lines resulted in a low signal to noise continuum, making it difficult to claim the detection of other, weaker emission lines.  However, the redshift calculated from Ly$\alpha$ does seem consistent with possible detections of other common emission features such as NV and Si IV.  The wavelength coverage of the spectrum was not large enough to detect Ly$\beta$ or OVI.    

We measured the rest-frame equivalent width (EW) and full width at half maximum (FWHM) of Ly$\alpha$ for SDSS J2228+0110.  The  wavelength coverage for this spectrum was too small to fit the continuum slope; instead, we assumed it to be a power law with slope $\alpha$ = -0.5 (f$_{\nu} \propto {\nu}^{\alpha}$).  The normalization of the power law was fit through a chi-squared minimization to the continuum redward of 1280$\AA$.  We fit the Ly$\alpha$ profile with a Gaussian on its red side, and we assumed the line to be symmetric to account for the Ly$\alpha$ forest on the blue side.  We find a rest-frame FWHM of 7.68 $\AA$, which is equivalent to a velocity of 1,890 km/s, making it a narrow Ly$\alpha$ but not abnormal.  The rest-frame EW is 21.9 $\AA$ and is at the low end, but well within the range of Ly$\alpha$ strengths of other z $\sim$ 6 quasars.

\section{Discussion}

High-redshift quasar searches have been popular over the last decade.  Large scale efforts have revealed that these objects are quite rare and quite difficult to find (\citealp{2000AJ....120.1167F}; \citealp{2007AJ....134.2435W}; \citealp{2008AJ....135.1057J}).  Many of these surveys use optical data to select candidates and rely on very red optical colors.  Unfortunately, high-z quasars share the same red optical color space with cool dwarf stars.  In an effort to reduce contamination, a blue near-IR cut is used in the selection method to help separate the two populations.  This technique successfully increases efficiency; however, it reduces the completeness of the quasar sample by an unknown amount.   Our incompleteness was estimated by redshifting low-z quasar spectra to high-z and tracking their colors.  This track indicates that a large ``blue'' quasar population exists at high-z, and that optical surveys are selecting a significant sample of the population.  However, it is an open question as to whether low-z quasars are truly like their high-z counterparts.  

\citet{2006ApJ...652..157M} discovered, at the time, the highest redshift radio-loud quasar (FIRST J1427+3312) using radio selection in a search area of only 4 deg$^2$.  The quasar was bright enough in the optical to be detected in other surveys, but its red near-IR color would have prevented its selection.  This rare discovery in such a small area hints that there may be a larger population of ``red'' quasars than predicted.  If this is true, current estimates for the quasar number density at z $\sim$ 6 are too low.  
 
Radio-selection offers an alternative method for the selection of high-z quasars.  Requiring a radio detection is just as efficient in removing contaminants as a blue near-IR color cut; however, radio-selection also allows for the detection of ``red'' quasars, possibly from dust reddening.  A likely cause of dust reddening is the circumnuclear cocoon that is thought to encase relatively young quasars (\citealp{ 2005ApJ...625L..71H}).  As the quasar ages, the dust cocoon may be blown away by energy released from accretion at near-Eddington luminosities.  The universe is less than one billion years old at z $\sim$ 6, and it would not be unreasonable that the fraction of quasars in this dust cocoon phase is higher at higher redshift.  
 
Of the seven z $\sim$ 6 quasars with $z-J >$ 0.8, three of them are radio-loud.  It is not unusual that radio-loud quasars are redder than their radio-quiet counterparts, as this effect is also seen at lower redshift (\citealp{2003AJ....126..706W}, \citealp{2009AJ....138.1925M}).  However, when quasars at lower redshift, z $\sim$ 3, are redshifted to z $=$ 6 and placed in the same colorspace as high-z quasars, at higher redshift radio-loud quasars tend to have redder $z-J$ colors than at lower redshift (see Figure 5).  This is a very intriguing result albeit in the small number regime, as it might suggest that radio-loud quasars are intrinsically redder at higher redshift or have higher quantities of dust.  The result should also be taken with caution as there is a substantial selection effect for z $\sim$ 3 quasars in SDSS (\citealp{richards}).  The selection efficiency at z $\sim$ 3 is $\sim$50$\%$ and biased towards redder $u-g$ colors; however, radio-selected z $\sim$ 3 quasars don't seem to show the same color bias (\citealp{worseck}).  This comparison of radio-loud/radio-selected z $\sim$ 3 quasars redshifted to z $=$ 6 with observed radio-loud z $\sim$ 6 quasars should be absent of a color selection-effect and serve as a viable comparison in cosmic time.    

Although dust reddening of the quasar continuum may contribute to a significant ``red'' quasar population, it would also cause high levels of extinction in the detection bands.  This extinction can be greater than three magnitudes for E(B-V) = 0.1 using a typical SMC  reddening law (\citealp{prevot}), placing even some of the brightest z $\sim$ 6 quasars below the detection limit of wide-area surveys like CFHT and SDSS.  A more likely cause of a significant ``red'' quasar population is the strength of Ly$\alpha$.  A strong Ly$\alpha$ line leads to a ``bluer'' population of high-z quasars in the $z-J$ color while a weak Ly$\alpha$ line leads to a ``redder'' population of z $\sim$ 6 quasars.  A second factor that can lead to a ``red'' quasar population is the absorption blueward of Ly$\alpha$ from neutral hydrogen in the intergalactic medium, which can lead to a ``red'' $i-z$ color for stronger absorption (see Figure 6).

This is still an ongoing search for radio-selected quasars.  We plan to observe more candidates in future observations and obtain a deeper optical spectrum of SDSS J2228+0110.  In an effort to measure the $z-J$ color of radio-selected z $\sim$ 6 quasars, we also plan to follow up on our current discovery and any new discoveries with near-IR imaging.  It would be interesting to use the discovered z $\sim$ 6 quasar to constrain the luminosity function of high-z quasars outside of the typical optical selection criteria; however, our survey is not complete.   
 
The greatest limitation to radio-selection is the depth of the radio data.  Only $\sim$5\% of z $\sim$ 6 quasars are detected in FIRST.  Even a medium depth radio survey like the Stripe82 VLA Survey does not reach the median quasar radio luminosity.  With a deep and wide-area radio survey, questions about the existence of a large ``red'' high-z quasar population, as well as the mechanism that might cause the reddening, could finally be answered.                  

{\bf Acknowledgements}:  Some of the data presented herein were obtained at the W.M. Keck Observatory, which is operated as a scientific partnership among the California Institute of Technology, the University of California and the National Aeronautics and Space Administration. The Observatory was made possible by the generous financial support of the W.M. Keck Foundation.
Based on observations obtained with MegaPrime/MegaCam, a joint project of CFHT and CEA/DAPNIA, at the Canada-France-Hawaii Telescope (CFHT) which is operated by the National Research Council (NRC) of Canada, the Institut National des Science de l'Univers of the Centre National de la Recherche Scientifique (CNRS) of France, and the University of Hawaii. This work is based in part on data products produced at TERAPIX and the Canadian Astronomy Data Centre as part of the Canada-France-Hawaii Telescope Legacy Survey, a collaborative project of NRC and CNRS. 
The authors wish to recognize and acknowledge the very significant cultural role and reverence that the summit of Mauna Kea has always had within the indigenous Hawaiian community.  We are most fortunate to have the opportunity to conduct observations from this mountain.  The National Radio Astronomy Observatory is a facility of the National Science Foundation operated under cooperative agreement by Associated Universities, Inc.
GRZ acknowledges NRAO Grant GSSP 09-0010.  The work by RHB was partly performed under the auspices of the U.S. Department of Energy by Lawrence Livermore National Laboratory under Contract DE-AC52-07NA27344.

\bibliographystyle{apj}
\bibliography{mybib}

\clearpage

\begin{deluxetable}{llcccccc}
\tabletypesize{\scriptsize}
\tablecaption{High Redshift Quasar Candidates Observed At Keck}
\tablewidth{0pt}
\tablehead{
\colhead{Number} & \colhead{Name} & \colhead{Exposure Time (s)} & \colhead{$z$} & \colhead{$z_{error}$} & \colhead{$i-z$} & \colhead{${(i-z)}_{error}$} & \colhead{$S_{1.4GHz,peak}$ (mJy)}
}
\startdata
 1 &  SDSS J004035.8+004900.6  &	900 & 21.42 & 0.08 & 1.85 & 0.21 &1.88\\
 2 &  SDSS J004942.8+003924.1  & 1800 & 22.14 & 0.13 & 1.96 & 0.27 & 1.53\\
 3 &  SDSS J005607.5-000401.4  & 1800 & 22.04 & 0.14 & 1.91 & 0.28 & 0.44\\
 4 &  SDSS J010006.7-004740.3 & 1800 & 21.98 & 0.15 & 1.98 & 0.34 & 0.52\\
 5 & SDSS J010140.3-001258.8  & 1800 & 22.24 & 0.20 & 4.90 & 2.04 & 0.35\\
 6 & SDSS J010543.6+004806.8 & 1800 & 22.24 & 0.19 & 1.76 & 0.35 & 0.42\\
 7 & SDSS J012406.0-005640.1 & 1800 & 22.24 & 0.20 & 3.74 & 4.01 & 2.90\\
 8 & SDSS J013453.8-004151.7 & 1800 & 22.15 & 0.15 &1.73 & 0.26 & 0.47\\
 9 & SDSS J013600.7-010010.2  & 1800 & 22.01 & 0.12 & 2.18 & 0.27 & 0.47\\
 10 & SDSS J014534.5-005641.6 & 1800 & 22.42 & 0.17 & 1.79 & 0.30 & 5.44\\
 11 & SDSS J015128.9-002037.5 & 1800 & 22.42 & 0.19 & 1.85 & 0.32 & 4.00\\
 12 & SDSS J020430.2+001834.3 & 1800 & 20.72 & 0.04 & 2.20 & 0.19 & 0.30\\
 13 & CFHT 021600.1-073546.4 & 1800 & 22.98 & 0.17 & 2.23 & 0.52 & 3.89\\
 14 & CFHT 021759.0-043256.5 & 1800 & 22.84 & 0.11 & 2.91 & 0.67 & 1.46\\
 15 & SDSS J021838.5+000557.4 & 1800 & 22.23 & 0.19 & 1.71 & 0.27 & 5.98\\
 16 & CFHT 022117.7-073947.3 & 1800 & 23.00 & 0.19 & 2.23 & 0.57 & 1.04\\
 17 & CFHT 022321.2-091011.1 & 1800 & 23.45 & 0.26 & 2.38 & 0.71 & 5.38\\
 18 & CFHT 022549.7-060052.1 & 1800 & 23.18 & 0.13 & 2.06 & 0.39 & 11.26\\
 19 & CFHT 085231.6-025004.2 & 1800 & 23.79 & 0.31 & 2.08 & 0.80 & 1.03\\
 20 & CFHT 085325.4-025913.3 & 2700 & 23.18 & 0.17 & 1.64 & 0.33 & 3.10\\
 21 & CFHT 085737.1-023530.0 & 1800 & 22.36 & 0.14 & 1.56 & 0.21 & 2.51\\
 22 & CFHT 090307.8-034327.9 & 3600 & 23.23 & 0.21 & 1.71 & 0.51 & 22.58\\
 23 & CFHT 090610.6-032423.1 & 1800 & 22.96 & 0.15 &1.77 & 0.31 & 29.39\\
 24 & CFHT 141227.8+531949.1 & 1800 & 23.00 & 0.17 & 2.24 & 0.53 & 1.92\\
 25 & CFHT 141240.9+514609.8 & 3600 & 23.35 & 0.17 & 1.65 & 0.33 & 8.21\\
 26 & CFHT 142105.5+554434.1 & 1800 & 23.54 & 0.24 & 2.26 & 0.82 & 11.22\\
 27 &  CFHT 142910.2+564216.6  & 900 & 22.69 & 0.17 & 2.62 & 0.55 & 4.20\\
 28 & CFHT 143414.2+523554.6 & 1800 & 23.37 & 0.23 &1.79 & 0.70 & 3.48\\
 29 & CFHT 221806.1+013619.8 & 1800 & 22.45 & 0.11 &1.82 & 0.25 & 2.86\\
 30 & SDSS J220755.9$-$005943.1 & 1800 & 22.19 & 0.18 &1.71 & 0.30 & 0.62\\
 31 & SDSS J222036.1$-$010119.9 & 900 & 22.12 & 0.18 & 2.13 & 0.41 & 0.65\\
 32 & SDSS J222843.5+011032.2\tablenotemark{a} & 3600 & 22.28 & 0.20 & 1.81 & 0.35 & 0.31\\
 33 & SDSS J223747.5+002521.0 & 1800 & 22.38 & 0.19 & 2.01 & 0.40 & 0.63\\
 34 & SDSS J223912.4$-$002803.0 & 1800 & 22.13 & 0.17 & 1.77 & 0.27 & 4.75\\
 35 & SDSS J231225.8$-$002020.0 & 900 & 22.32 & 0.20 & 1.84 & 0.36 & 0.78\\ 

\enddata
\tablenotetext{a}{Quasar found in this paper, z $=$ 5.95}

\end{deluxetable}

\clearpage
\begin{figure}[htp]
\centering
\includegraphics[totalheight=0.5\textheight,width=1\textwidth]{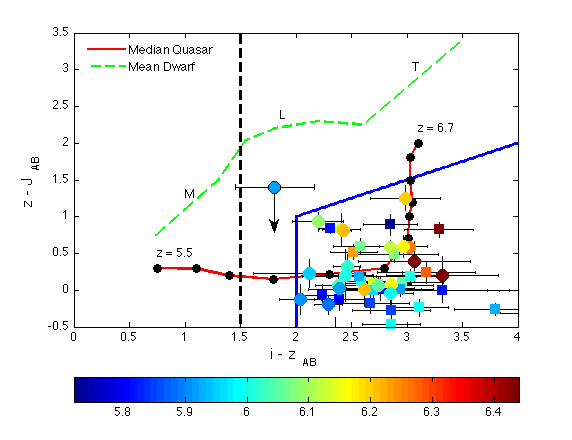}
\caption{Color-Color diagram illustrating the photometric quasar selection of \citet{2009AJ....137.3541W}, which is bounded by the two solid lines, and that of this project, which is a vertical dashed line at $(i-z)_{AB}$ $>$ 1.5 for CFHT ($(i-z)_{AB}$ $>$ 1.7 for SDSS).  The red solid line is a median composite of template quasars redshifted from z = 5.5 to z = 6.7, and the dashed jagged green line is the mean location of M, L, and T dwarfs (\citealp{2009AJ....137.3541W}).  The quasar discovered in this paper has upper limit in the $z-J$ direction as it was as it was not detected in UKIDSS in the $J$ band.  The colors represent redshift in which blue is z=5.7 and red is z=6.5.  The filled squares are quasars discovered in CFHT, and the filled circles are quasars discovered in SDSS or other surveys}\label{fig:f1}
\end{figure}

\clearpage
\begin{figure}[htp]
\centering
\includegraphics[totalheight=0.4\textheight,width=.45\textwidth]{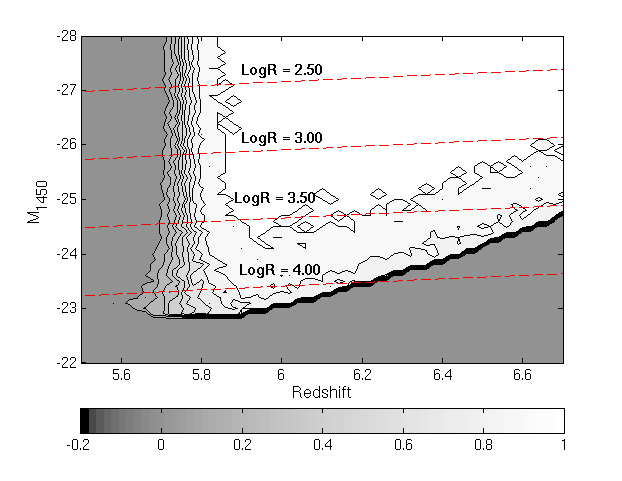}
\includegraphics[totalheight=0.4\textheight,width=.45\textwidth]{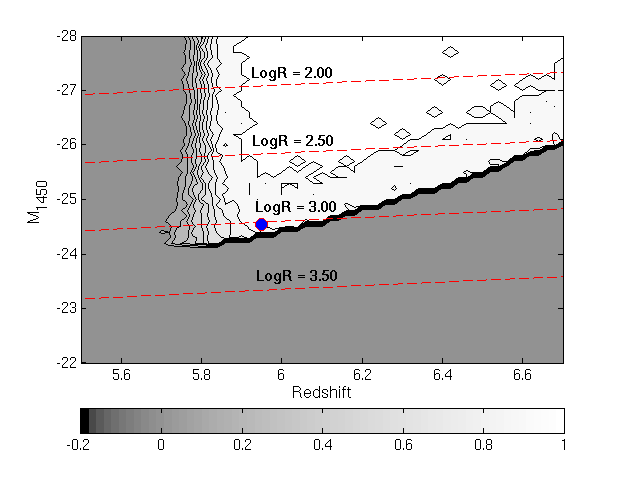}
\caption{Completeness as a function of redshift and absolute magnitude, $M_{1450}$.  The white equals 100\% complete and the dark gray equals 0\% complete and the contours are in linear steps of 10$\%$.  The figure on the left is CFHTLS Wide matched to FIRST.  The red dashed lines represent levels of radio loudness, LogR $=$ Log ($f_{5GHz}/f_{2500}$), which is calculated assuming a power-law of $f_{\nu} \propto \nu^{\alpha}$ (where $\alpha = -0.5$ for both).  The figure on the right is Stripe 82 matched to the Stripe82 VLA Survey.  The blue circle is the quasar discovered in this paper.}\label{fig:f1}
\end{figure}

 \clearpage
\begin{figure}[htp]
\centering
\includegraphics[totalheight=0.25\textheight,width=1\textwidth]{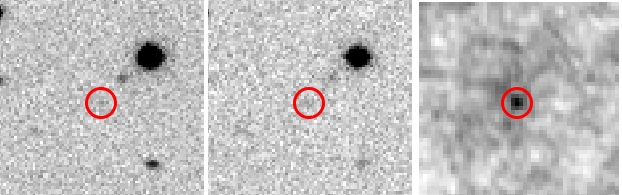}
\caption{30'' cutout images of Stripe 82 $i$, $z$, and VLA 1.4 GHz for SDSS J2228+0110.  The Stripe82 VLA position is marked with a red circle.  The quasar was at the detection limit for all three bands.}\label{fig:f1}
\end{figure}

 \clearpage
\begin{figure}[htp]
\centering
\includegraphics[totalheight=0.6\textheight,width=1\textwidth]{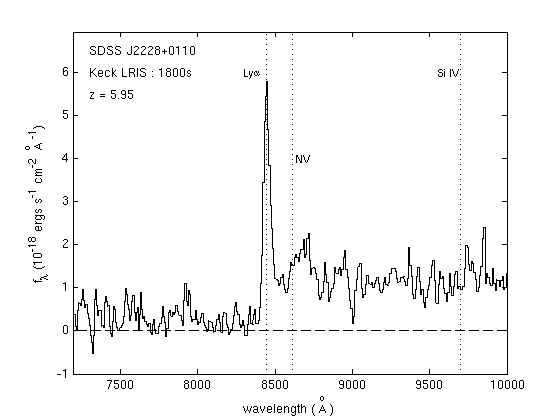}
\caption{Spectrum of SDSS J222843.54+011032.2 obtained in June 2010 using LRIS on the Keck 1 telescope.  The spectrum has been binned using inverse sky-variance weighting to reduce the sky noise.  Only the two best out of the total four exposures were used to make this optimal spectrum.  The redshift, z $=$ 5.95, was calculated using the peak of the identified Ly$\alpha$ emission line.  The expected locations of other typical emission lines are labeled with a vertical dotted line.}\label{fig:f1}
\end{figure}

 \clearpage
\begin{figure}[htp]
\centering
\includegraphics[totalheight=0.6\textheight,width=1\textwidth]{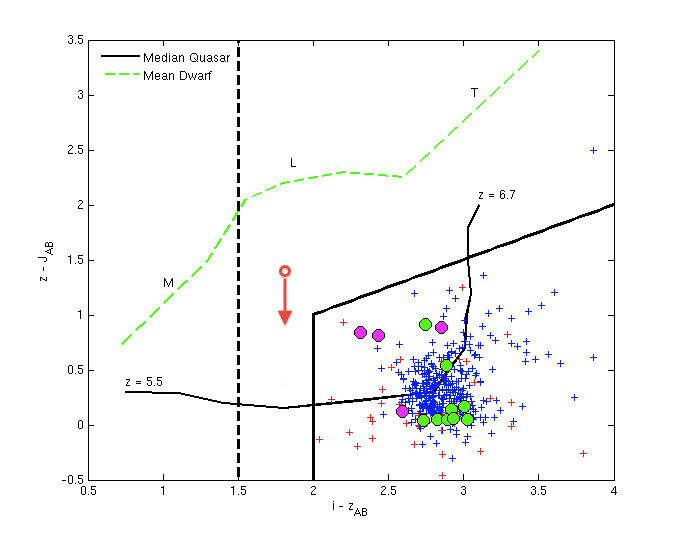}
\caption{Color space comparison of z $\sim$ 6 quasars and z $\sim$ 3 quasars redshifted to z $=$ 6.  The lower redshift quasar sample consists of 372 quasars with 2.98 $<$ z $<$ 3.02 and was taken from the SDSS Data Release 7 (DR7) database.  The SDSS spectra of these quasars were used to calculate the $i$ - $z$ and $z$ - $J$ colors.  The spectra were corrected for Ly$\alpha$ and Ly$\beta$ absorption from the IGM according to \citet{2006AJ....132..117F}.  Red crosses are z $\sim$ 6 quasars which were not detected at 1.4 GHz above 0.1 mJy, while magenta circles are z $\sim$ 6 quasars which were detected.  Blue crosses are z $=$ 3 quasars which were not detected at 1.4 GHz above 1 mJy, and green circles are z $=$ 3 quasars which were detected.  Of the nine z $=$ 3 quasars which were detected in FIRST, seven of them were radio-selected while the other two were color-selected with a radio detection.  The two that were color-selected have $z-J=0.06$ and $z-J=0.91$. The red circle is the quasar found in this paper and only has an upper limit for its $z - J$ color.  The radio-loud z $\sim$ 6 quasars (mean $z-J$ $=$ 0.67) are significantly redder than the radio-loud, z $=$ 3 quasars (mean $z-J$ $=$ 0.18).  The black solid line is a median composite of template quasars redshifted from z = 5.5 to z = 6.7 and the dashed jagged green line is the mean location of M, L, and T dwarfs (\citealp{2009AJ....137.3541W}). }
\end{figure}

 \clearpage
\begin{figure}[htp]
\centering
\includegraphics[totalheight=0.6\textheight,width=1\textwidth]{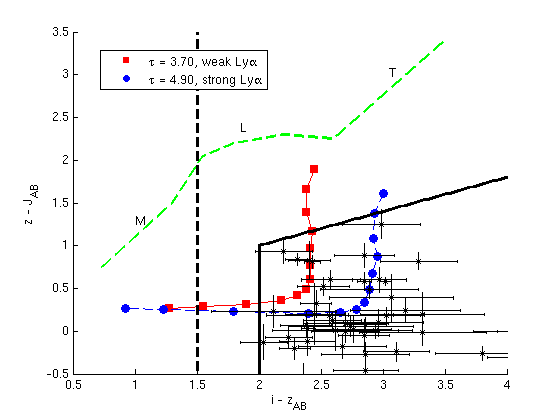}
\caption{This plot demonstrates the effect of neutral hydrogen absorption and the strength of Ly$\alpha$ on the high-z quasar track.  A composite quasar from \citet{2001AJ....122..549V} was redshifted from z$=$5.5 to z$=$6.7 using an effective Gunn-Peterson absorption, $\tau$, due to Ly$\alpha$ and Ly$\beta$ absorption.  The strength of the Ly$\alpha$ emission was also allowed to vary.  The black box is the typical optical selection used by SDSS and CFHT.  The green dashed line is the mean dwarf star track, while the vertical black dashed line is the optical selection used in this paper for CFHTLS Wide.  The black points are known z $\sim$ 6 quasars with their associated photometric errors.  The red track is the composite quasar redshifted from z$=$5.5 to z$=$6.7 with an effective Gunn-Peterson optical depth of $\tau$ = 3.70 and a weak Ly$\alpha$ emission line.  The blue track is the composite quasar redshifted from z$=$5.5 to z$=$6.7 with an effective Gunn-Peterson optical depth of $\tau$ = 4.90 and a strong Ly$\alpha$ emission line.  A higher optical depth causes redder $i-z$ colors while a weaker Ly$\alpha$ emission causes redder $z-J$ colors. }\label{fig:f1}
\end{figure}

\end{document}